\documentclass[twocolumn]{aastex63}

\usepackage{graphicx}	
\usepackage{amsmath}	
\usepackage{amssymb}	
\usepackage[shortlabels]{enumitem}      
\usepackage{booktabs}   

\definecolor{orange}{rgb}{0.8, 0.35, 0}
\definecolor{green}{rgb}{0.3, 0.6, 0.3}

\graphicspath{{Figures}}
\DeclareGraphicsExtensions{.png,.pdf,.eps}

\received{}
\revised{}
\accepted{}
\submitjournal{ApJ}

\shorttitle{DM annihilation during Cosmic Dawn}
\shortauthors{List et al.}

\begin{document}

\title{Lux ex tenebris: The imprint of annihilating dark matter on the intergalactic medium during Cosmic Dawn}

\author[0000-0002-3741-179X]{Florian List}
\affiliation{Sydney Institute for Astronomy, School of Physics, A28, The University of Sydney, NSW 2006, Australia}

\author[0000-0002-6154-7224]{Pascal J. Elahi}
\affiliation{International Centre for Radio Astronomy Research, University of Western Australia, 35 Stirling Highway, Crawley, WA 6009, Australia}
\affiliation{ARC Centre of Excellence for All Sky Astrophysics in 3 Dimensions (ASTRO 3D)}

\author[0000-0003-3081-9319]{Geraint F. Lewis}
\affiliation{Sydney Institute for Astronomy, School of Physics, A28, The University of Sydney, NSW 2006, Australia}

\correspondingauthor{F. List}
\email{florian.list@sydney.edu.au}



\begin{abstract}

Upcoming measurements of the highly redshifted 21cm line with next-generation radio telescopes such as HERA and SKA will provide the intriguing opportunity to probe dark matter (DM) physics during the Epoch of Reionization (EoR), Cosmic Dawn, and the Dark Ages. 
With HERA already under construction, there is a pressing need to thoroughly understand the impact of DM physics on the intergalactic medium (IGM) during these epochs. We present first results of a hydrodynamic simulation suite with $2 \times 512^3$ particles in a $(100 \ h^{-1} \ \text{Mpc})^3$ box with DM annihilation and baryonic cooling physics. We focus on redshift $z \sim 11$, just before reionization starts in our simulations, and discuss the imprint of DM annihilation on the IGM and on structure formation. We find that whereas structure formation is not affected by thermal WIMPs heavier than $m_\chi \gtrsim 100 \ \text{MeV}$, heating from $\mathcal{O}$(GeV) DM particles may leave a significant imprint on the IGM that alters the 21cm signal. Cold gas in low density regions is particularly susceptible to the effects of DM heating. We note, however, that delayed energy deposition is not currently accounted for in our simulations. 

\end{abstract}



\section{Introduction}
Despite strenuous efforts by the scientific community, dark matter (DM) still successfully eludes its detection by state-of-the-art particle colliders such as the LHC \citep{Abdallah2015} and underground detectors \citep{Schumann2019}. For decades, weakly interacting massive particles (WIMPs) have been amongst the most popular DM candidates \citep{Bertone2005}. These particles were in a thermal equilibrium with the baryonic plasma in the very early Universe and froze out as the Universe expanded. In order to explain the observed relic DM density, WIMPs should have an annihilation cross section at the weak scale where many hypothetical particles such as the neutralino or the lightest Kaluza--Klein particle reside -- a coincidence commonly dubbed the ``WIMP miracle''. However, constraints on the WIMP mass $m_\chi$ and the annihilation velocity cross section $\langle \sigma v \rangle$ from the CMB \citep{PlanckCollaboration2018}, from $\gamma$-ray measurements \citep{Ackermann2015, Albert2017}, and from cosmic ray observations \citep{Cuoco2017} have excluded large parts of the $m_\chi - \langle \sigma v \rangle$ plane by now (see \citealt{Leane2018} for a combined analysis that derives the lower bound $m_\chi \gtrsim 20 \ \text{GeV}$ for thermal WIMPs annihilating via s-wave $2 \to 2$ annihilation into visible final states). 
\par An exciting prospect is the launch of the upcoming radio interferometers HERA \citep[\emph{Hydrogen Epoch of Reionization Array;}][]{DeBoer2017} and SKA \citep[\emph{Square Kilometer Array;}][]{Mellema2013} that will allow probing the dark sector in the uncharted redshift range between the most distant quasars observed to date at $z \gtrsim 7$ \citep{Banados2018} and the CMB at $z \sim 1100$. In particular, the spin-flip 21cm line from neutral hydrogen, received redshifted in the radio frequencies, is expected to reveal unprecedented insights into the Dark Ages and the Epoch of Reionization \citep[EoR, e.g.][]{Furlanetto2006, Pritchard2012}. Annihilating DM particles produce standard model particles that deposit a fraction of their energy into their surroundings, thereby heating and ionizing the intergalactic medium (IGM). Although the contribution of the resulting energy production to the increase of the free electron fraction $x_e$ is small, e.g. $\lesssim 10 \%$ for s-wave and p-wave annihilation via the channels $\chi \chi \to e^+ e^-$ and $\chi \chi \to \gamma \gamma$ \citep{Liu2016}, and is expected to play at most a secondary role in reionizing the Universe \citep{Lopez-Honorez2013, Poulin2015}, DM heating may leave a detectable imprint on the global 21cm signal and power spectrum \citep{Furlanetto2006a}. 
During the Dark Ages, physics are relatively simple due to the absence of astrophysical sources, thus providing a pristine testing ground for exotic physics such as DM annihilation. In contrast, there is a large uncertainty about the photon sources triggering reionization, and distilling the effects of annihilating DM on the 21cm line during the EoR poses a difficult task. While \citet{Valdes2013} find that a 10 GeV bino-like particle strongly affects the global 21cm brightness temperature and \citet{Evoli2014} report that a 10 GeV leptophilic thermal WIMP annihilating via $\chi \chi \to \mu^+ \mu^-$ can be detected by HERA and SKA, \citet{Lopez-Honorez2016} caution that particles as light as $m_\chi \sim 100$ MeV are necessary for unequivocally ascertaining a signature from DM annihilation in the 21cm signal.
\par Studying the effects of DM annihilation during the Dark Ages and the EoR often relies on analytic models \citep[e.g.][]{Chuzhoy2008}, modifications of recombination codes such as \textsc{Recfast} \citep{Seager1999} as in \citet{Cumberbatch2010, Oldengott2016}, or adaptations of semi-numerical schemes such as \textsc{21cmFAST} \citep{Mesinger2011} as done in e.g. \citet{Evoli2014}. The evolution of the density field is commonly modeled with the Zel'dovich approximation \citep{zel1970gravitational}, formalisms are invoked for the halo mass function \citep[HMF, e.g.][]{press1974formation, Sheth1999}, and Navarro--Frenk--White halo profiles are assumed \citep{Navarro1997}. However, in order to spatially resolve the complex interplay between heating and ionization from DM, baryonic cooling physics, hydrodynamics, and gravity over the course of cosmic time, hydrodynamic simulations are an indispensable tool. 
\par In this work, we present the results of a suite of hydrodynamic simulations until the end of Cosmic Dawn with $2 \times 512^3$ particles in a $(100 \ h^{-1} \ \text{Mpc})^3$ box that incorporate DM annihilation and cooling physics. DM particles in the simulations deposit energy into the surrounding gas at each time step, the amount depending on the local dark matter density and the respective DM candidate. Therefore, no analytic models for the HMF or halo profiles are required, and the heat generated by DM annihilation directly affects the evolution of the gas -- unlike when calculating the strength of DM annihilation in a post-processing step. Throughout this paper, we focus on redshift $z \sim 11$, shortly before UV radiation from baryonic sources reionizes the Universe.
We study this epoch for two reasons: first, at a redshift of $z \sim 11$ (corresponding to a received 21cm frequency of $\nu \sim 118 \ \text{MHz}$), already the Baseline Design SKA1-Low \citep{Dewdney2013} will be able to carry out 21cm imaging at low noise levels, while the image quality rapidly deteriorates for $\nu \lesssim 100 \ \text{MHz}$, i.e. when peering deeper into Cosmic Dawn \citep{Mellema2015}; and second, considering the end of Cosmic Dawn allows us to assume that the Wouthuysen--Field effect is fully saturated \citep{Wouthuysen1952, Field1958}, irrespective of the large uncertainty about the emission from the first astrophysical sources. 
\par We proceed as follows: 
in Section \ref{sec:sims}, we briefly introduce our implementation of DM annihilation into the hydrodynamic simulation code \textsc{Gizmo}; moreover, we summarize the simulation parameters, the modeling of baryonic physics, and the investigated DM candidates.
In Section \ref{sec:results}, we analyze the density and temperature distribution of the hydrogen gas; furthermore, we consider the implications of DM heating on the HMF. Finally, we show the impact of DM annihilation on the 21cm line.
We discuss our findings in Section \ref{sec:discussion} and examine the validity of the 
common simplification that assumes spatially homogeneous DM heating, enhanced due to substructure by a redshift-dependent boost factor $\mathcal{B}(z)$, in the calculation of the 21cm brightness temperature.

\section{Simulations}
\label{sec:sims}
For incorporating DM annihilation, we developed a module for the hydrodynamic simulation code \textsc{Gizmo} \citep{GIZMO}, which is an offshoot of the popular \textsc{Gadget} series \citep{Gadget2}.
The simulations track the evolution of $512^3$ N-body gas and DM particles of masses $1.48$ and $7.95 \times 10^8 \ \text{M}_\odot$, respectively, in a periodic $(100 \ h^{-1} \ \text{cMpc})^3$ box. The softening length is taken to be $9.77 \ h^{-1} \ \text{ckpc}$, and the effective neighbor number for the reconstruction of hydrodynamic quantities and for the deposition of DM energy into the surrounding gas is 40. We assume \citet{Ade2016} cosmology 
and create the initial conditions at $z = 100$ with the tool \textsc{NGenic} \citep{NGenic}, neglecting DM annihilation prior to that redshift. For solving the Euler equations, we employ the meshless finite mass method. Radiative cooling is enabled that accounts for heating and cooling from H and He ionization and recombination, collisional, free-free, and Compton effects, molecular cooling down to 10 K, and metal line cooling for 11 species \citep{FIRE2, Wiersma2009}, so as to be maximally conservative with regards to the various cooling mechanisms that counteract the DM heating. Note that since the cooling times are generally much shorter than the dynamical times for gas temperatures $\lesssim 10^6 \ \text{K}$, galaxy formation has been found to depend only very weakly on the exact modeling of cooling physics \citep{FIRE2}. Star formation follows the description in \citet{Springel2003}, and we do not model any stellar feedback in order to extract the signature of DM annihilation. 
\par For the treatment of DM annihilation, we use the ``donor-based'' method presented in \citet{List2019}, in which the energy production is evaluated at each N-body DM particle (in contrast to the ``receiver-based'' method proposed in \citealt{Iwanus2017}, where the DM heating is computed at each N-body gas particle). The generated energy is distributed among the neighboring gas particles in an (approximately) isotropic way, resulting in an instantaneous and localized energy deposition. 
It has been shown that delayed energy deposition suppresses the heat absorbed by the IGM for annihilation \citep{Slatyer2016,DAmico2018}. However, accounting for this effect in N-body simulations would require the use of tracer particles and introduce new modeling choices for their interaction with gas particles, magnetic fields, etc., for which reason we leave this to future work.
The energy produced by each N-body DM particle in the simulation is given by
\begin{equation}
\label{eq:DMAF}
\frac{\text{d}E}{\text{d}t} = \mathcal{B} f \frac{\langle\sigma v\rangle}{m_\chi} \rho_{\chi} M c^2,
\end{equation}
where $\mathcal{B}$ is a boost factor accounting for unresolved substructure, $f$ is the energy absorption fraction of the IGM, $\rho_\chi$ is the local DM density, and $M$ is the mass of the annihilating N-body DM particle. Since the mass loss due to DM annihilation is negligible except for extremely light DM particles, we keep the masses of the N-body DM particles constant in the simulations. We assume constant $\langle\sigma v\rangle$ as is the case for s-wave annihilation. For presentation purposes, we set $\mathcal{B} = f = 1$ and $\langle \sigma v \rangle = 3 \times 10^{-26} \ \text{cm}^3 \ \text{s}^{-1}$ at the thermal relic value \citep{Steigman2012} and view $m_\chi$ as a varying parameter, but note that scaling the constituents of the effective parameter $p = \mathcal{B} f \langle \sigma v \rangle / m_\chi$ at fixed $p$ results in the same generated energy in our method; hence, each of the variables $\langle \sigma v \rangle$, $m_\chi$, $\mathcal{B}$, and $f$ can be viewed as a free parameter while keeping the others fixed. The entire energy from DM annihilation in our simulations is used for heating the gas and does not directly affect its ionization fraction. 
We note that the simplifying assumption of $f = 1$ leads to an overestimation of the amount of DM heating for a given DM particle mass; however, within the assumptions made in this work, namely instantaneous and redshift-independent energy absorption, a choice of e.g. $f = 0.1$ instead of $f = 1$ is equivalent to a ten times larger DM particle mass. This reinterpretation shall be carefully revisited in future work taking into account realistic energy absorption and computing the energy fractions going into gas heating, ionization, Ly-$\alpha$ photons, and free-streaming photons as a function of DM candidate, redshift, and ionization fraction \citep[see e.g.][]{Evoli2012, Slatyer2013, Slatyer2016, Liu2020}. 
We run simulations for $m_\chi \in \{1 \ \text{MeV}, \ldots, 100 \ \text{GeV}\}$ in logarithmic steps of 10, but we will focus mostly on the heavy mass range that is not in tension with current observations.

\section{Results}
\label{sec:results}
\subsection{Heating the cold IGM}
\begin{figure}
  \centering
  \noindent
  \resizebox{\columnwidth}{!}{
  \includegraphics{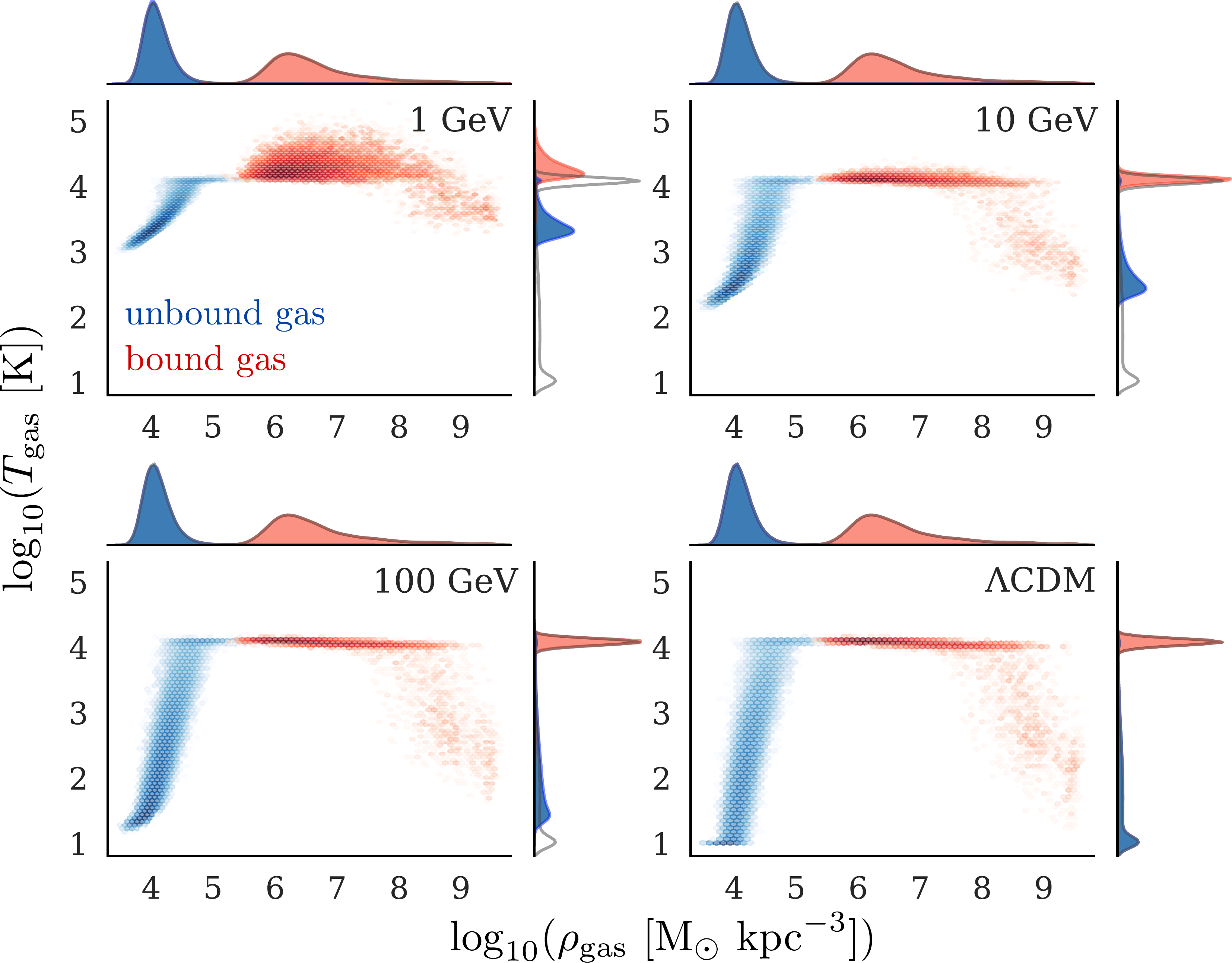}
  }
  \caption{Gas temperature $-$ density phase space distribution for different DM masses at $z = 11$, split up into gas particles bound in haloes (\emph{red}) and unbound gas particles (\emph{blue}). Grey lines in the marginal plots show the $\Lambda$CDM simulation without DM annihilation for reference.}

  \label{fig:T_vs_rho}
\end{figure}
Figure \ref{fig:T_vs_rho} shows the temperature -- density phase space distribution of gas bound in haloes (\emph{red}) and unbound gas (\emph{blue}) at $z = 11$. To identify haloes and the gas associated with them, we use  \textsc{VELOCIraptor} \citep{Elahi2019}. For all DM masses, there is little change in the density distribution -- even for light DM particles for which DM heating is significant. For the $1 \ \text{GeV}$ case, the minimum temperature of unbound gas particles increases by two orders of magnitude, resulting in the unbound gas only spanning roughly one order of magnitude in temperature as compared to three for the fiducial model without DM annihilation. Also for heavier DM candidates of masses $m_\chi = 10 - 100 \ \text{GeV}$, the impact of DM heating on the unbound gas particles is noticeable and still amounts to more than one order of magnitude in temperature for the cold gas in voids in the $10 \ \text{GeV}$ case.
\par As to gas in haloes, substantial heating is noticeable for $m_\chi = 1 \ \text{GeV}$, and $m_\chi \leq 10 \ \text{GeV}$ is still enough for DM heating to counteract the cooling of hot gas in very dense regions to temperatures $\sim 100 \ \text{K}$. 
Note that at $z = 11$, the mass fraction of bound gas particles amounts to only $0.012 \%$, and the hot dense gas would barely be visible in the phase space distribution plot if the distribution of \emph{all} the gas particles were plotted jointly. Thus, the primary effects of DM heating occur in regions outside of haloes, despite haloes being where the DM density is high and the feedback from DM annihilation is consequently the strongest.

\subsection{Imprint on structure formation}
\begin{figure}
  \centering
  \noindent
  \resizebox{\columnwidth}{!}{
  \includegraphics{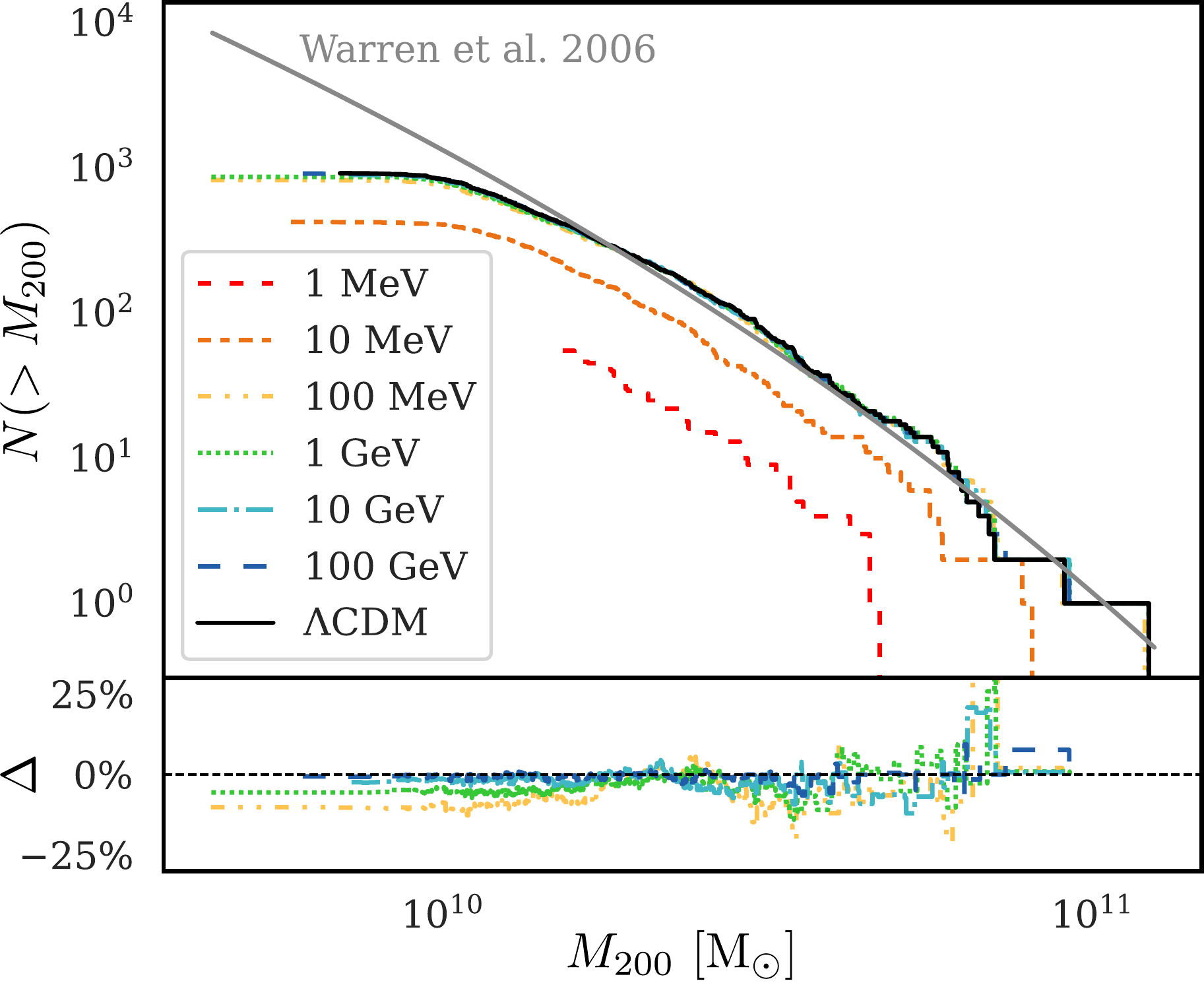}
  }
  \caption{Halo mass function for different DM masses at $z = 11$. The difference towards the reference $\Lambda$CDM simulation without DM annihilation is plotted below. The formation of haloes is largely unaffected by heating from DM particles with $m_\chi \geq 100 \ \text{MeV}$. The gray line shows the empirical fitting function proposed by \citet{Warren2006} for comparison.}
  \label{fig:hmf}
\end{figure}
In order to examine whether the heat generated by DM annihilation alters the formation of structure, we compute the HMF for different DM masses (see Figure \ref{fig:hmf}). We adopt the convention $\Delta = 200$ and define the virial mass with respect to the critical density $\rho_\text{crit}$. For DM masses $m_\chi \gtrsim 100 \ \text{MeV}$, the impact of DM annihilation on the formation of haloes is small. Only for very light DM candidates is the heating strong enough to drive the gas out of haloes, curbing halo formation by a significant amount. Whereas the HMF in \citet{Iwanus2019} at high redshift is reduced for a $100 \ \text{MeV}$ WIMP in a simulation without baryonic cooling physics, this effect is largely erased when taking cooling into account, and the reduction only amounts to $\sim 10\%$ for the smallest haloes. For haloes consisting of $N \gtrsim 40$ particles, the popular HMF model by \citet{Warren2006}, which lies between the Press--Schechter and Sheth--Tormen models, matches the reference simulation without DM annihilation well. 
\par We remark that for extremely light DM particles, star formation is severely impeded as the simulations run past reionization: for $m_\chi = 1 \ \text{MeV}$, barely any stars form by redshift $z \sim 3.5$, and the number of stars is reduced by a factor of two for $m_\chi = 10 \ \text{MeV}$ as compared to the reference simulation without DM annihilation. For $m_\chi \geq 1 \ \text{GeV}$, star formation activity is not affected by DM annihilation.

\subsection{Impact on the 21cm line}
\begin{figure*}
  \centering
  \noindent
  \resizebox{1\textwidth}{!}{
  \includegraphics{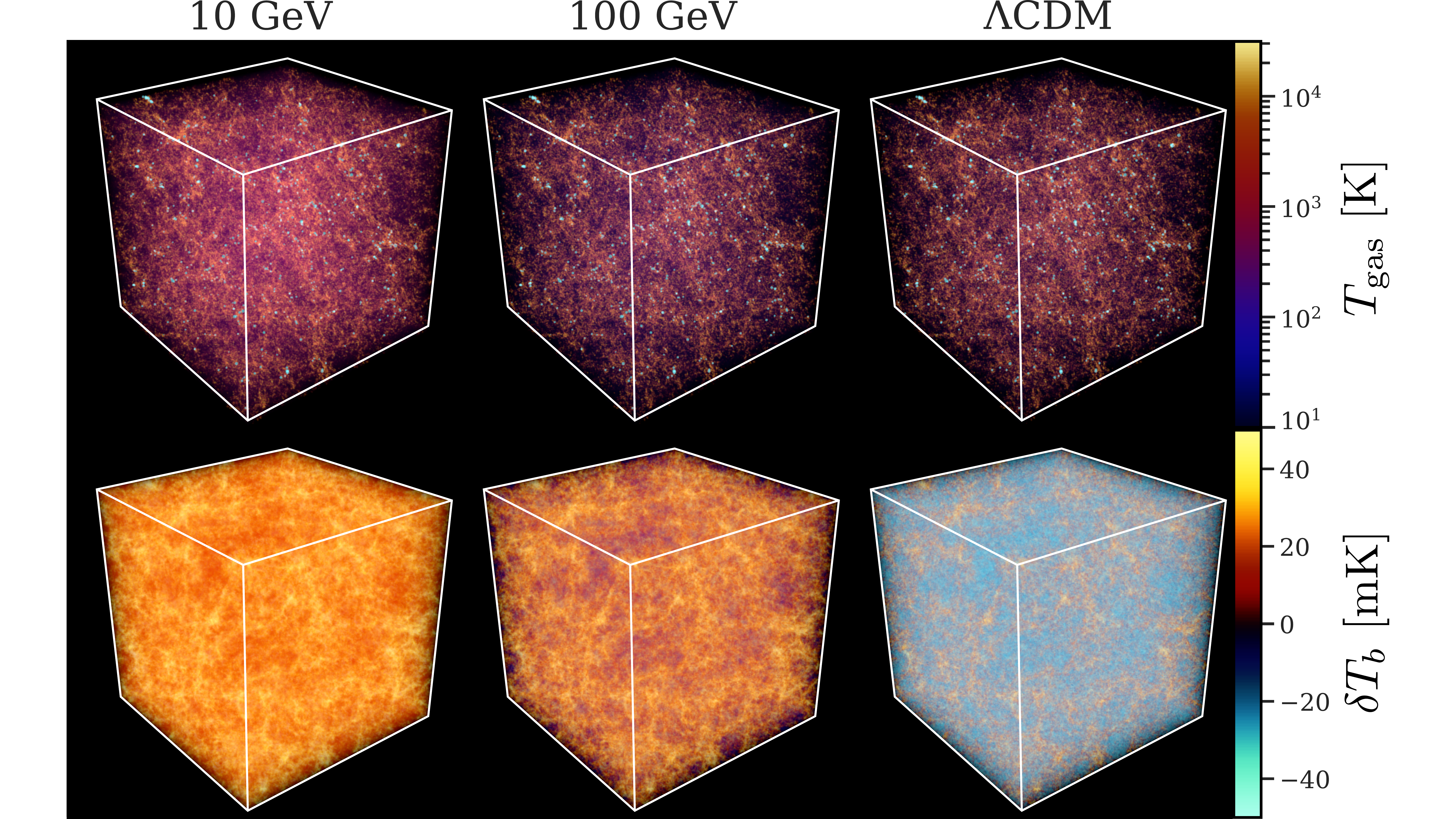}
  }
  \caption{Gas temperature (\emph{top}) and 21cm brightness temperature (\emph{bottom}) at $z = 11$. The light blue spheres are located at the centers of mass of dark matter haloes and their radii correspond to $5 \, R_\text{vir}$. For $m_\chi = 10 \ \text{GeV}$, the cold HI gas is universally heated, causing the 21cm line to be in emission everywhere. For a $m_\chi = 100 \ \text{GeV}$ particle, DM heating is strong enough to induce $\delta T_b > 0$ in vast parts of the filamentary structure, whereas $\delta T_b \lesssim 0$ in voids. Without DM annihilation, only gas in dense regions (and in particular in haloes) can be seen in emission. An animated version of this figure is available in the HTML version of this paper, which shows the simulation boxes from different angles and zooms into the boxes.}
  \label{fig:Tgas_and_Tb}
\end{figure*}
The differential brightness temperature $\delta T_b$ against the CMB is commonly written as
\begin{equation}
\begin{aligned}
    \delta T_b
&\approx 27 x_{\text{HI}}(1+\delta)\left(\frac{T_{s}-T_{\gamma}}{T_{s}}\right) \left(1+\frac{1}{H(z)} \frac{\text{d} v_{\|}}{\text{d} r_{\|}}\right)^{-1} \\  
&\quad \times \left(\frac{1+z}{10}\right)^{\frac{1}{2}}\left(\frac{\Omega_{b}}{0.044} \frac{h}{0.7}\right)\left(\frac{\Omega_{m}}{0.27}\right)^{\frac{1}{2}} \ \text{mK},
\end{aligned}
\end{equation}
where $x_{\text{HI}}$ is the local HI fraction, $T_s$ and $T_\gamma$ are the spin temperature and the CMB temperature, respectively, $\delta$ is the local fractional overdensity, and $\text{d} v_{\|} / \text{d} r_{\|}$ is the velocity gradient along the line of sight. Although peculiar velocities can increase the power spectrum by a factor of two \citep{Barkana2005}, we do not choose a line-of-sight direction herein and neglect the velocity gradient term in our calculations since we are only interested in comparing different models. Additionally, we can safely assume that the Wouthuysen--Field effect is saturated by $z \sim 11$ and therefore set $T_s = T_\text{gas}$. The local HI fraction $x_{\text{HI}}$ is computed by \textsc{Gizmo} for each N-body gas particle.
\par Figure \ref{fig:Tgas_and_Tb} shows the spatial distribution of the gas temperature $T_\text{gas}$ (\emph{top}) and the differential brightness temperature $\delta T_b$. (\emph{bottom}). 
The light blue spheres mark the locations of the haloes, and their radii are given by 5 times the virial radius $R_\text{vir}$. The main effect of DM annihilation is the heating of the cold HI gas in voids, which is reflected in the purple hue in the gas temperature plots. 
For a WIMP of $m_\chi = 10 \ \text{GeV}$, the DM annihilation causes the gas to surpass the CMB temperature in the entire simulation box, and the 21cm line is globally in emission. For the $100 \ \text{GeV}$ WIMP, only voids remain in absorption, while large regions are in emission. In contrast, the volume-averaged brightness temperature in the case without DM annihilation is negative, and only gas within filamentary structures is seen in emission.
\par
\begin{figure}
  \centering
  \noindent
  \resizebox{0.8\columnwidth}{!}{
  \includegraphics{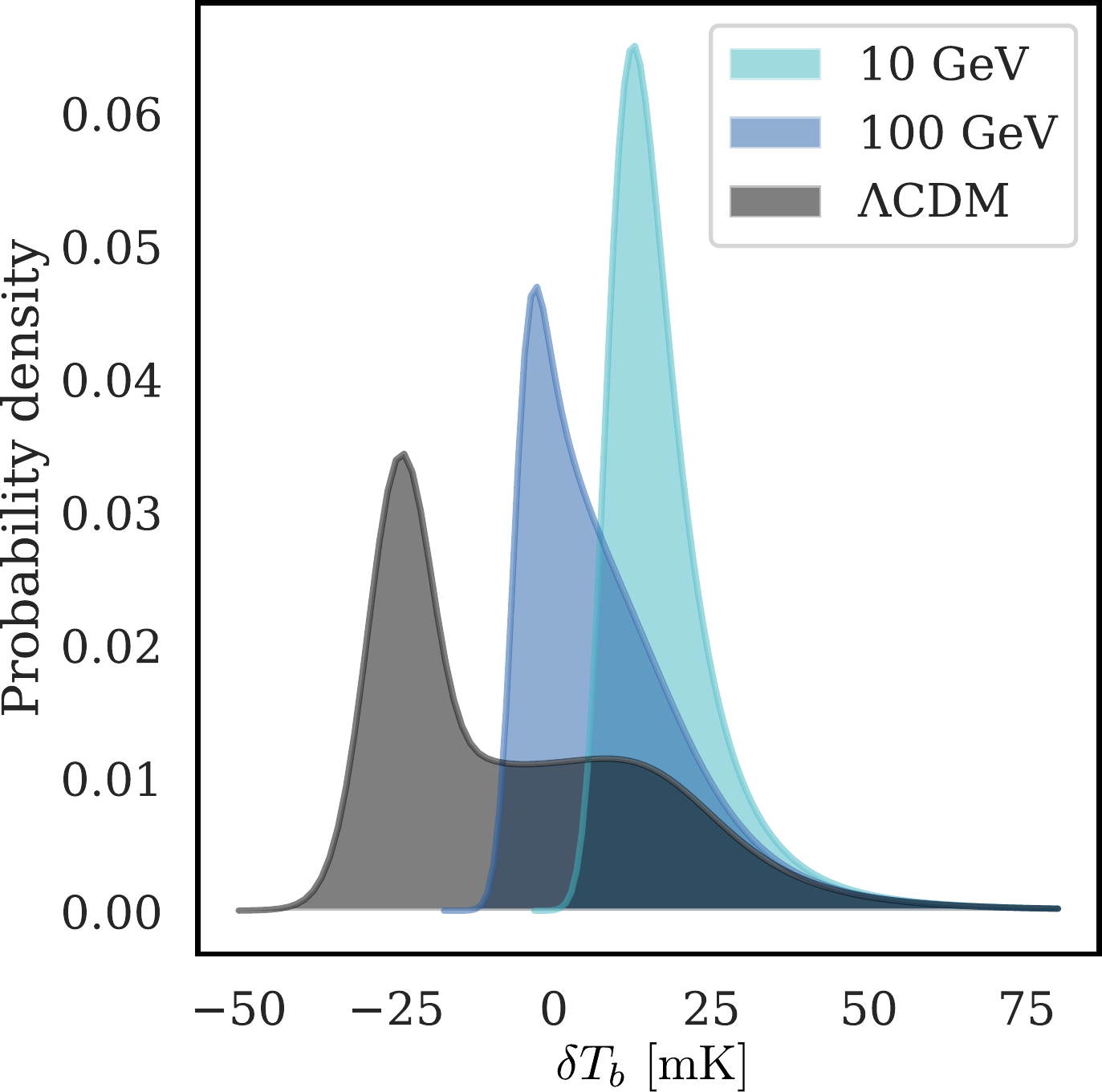}
  }
  \caption{Brightness temperature distribution at $z = 11$. For $m_\chi = 10 \ \text{GeV}$, the 21cm signal is globally in emission, whereas for $m_\chi = 100 \ \text{GeV}$ and without DM annihilation, roughly $2/3$ and $1/3$ of the simulation box are in emission, respectively.}
  \label{fig:Tb_hist}
\end{figure}

 This is confirmed by considering the distribution of brightness temperature values per cell, depicted in Figure \ref{fig:Tb_hist}. The distribution is computed by mapping the brightness temperature values from the N-body gas particles onto a regular grid of size $512^3$ using Shepard interpolation, binning the resulting values, and using a kernel density estimator. The distribution of $\delta T_b$ peaks at $\sim -24, -3,$ and $13 \ \text{mK}$ without DM annihilation, for $m_\chi = 100 \ \text{GeV}$, and for $m_\chi = 10 \ \text{GeV}$, respectively, while the means are located at $\sim -6, 9,$ and $18 \ \text{mK}$. The brightness temperature distribution becomes narrower as the DM mass decreases. This is in line with findings by \citet{Valdes2013, Evoli2014} who show that the 21cm power spectrum is reduced due to the relative uniformity of DM heating as compared to heating from astrophysical sources. In our simulations, a minimum temperature floor of $T_\text{gas} = 10 \ \text{K}$ is set for numerical stability, for which reason the lower end of the $\delta T_b$ distribution should not be over-interpreted. 

\section{Discussion}
\label{sec:discussion}
\par In this paper, we have presented first results from a suite of hydrodynamic simulations that self-consistently include annihilating DM in conjunction with baryonic cooling physics at high redshift. We have analyzed how the spatial distribution and the $T - \rho$ phase space distribution of the hydrogen gas at the end of Cosmic Dawn are affected by annihilating DM. Running hydrodynamic simulations instead of resorting to approximate methods allows us to evaluate the power generated by DM annihilation as a function of the non-linear local DM density field and to locally deposit the result DM heating, without the need for analytic halo mass functions or halo profiles. 
This is in contrast to previous investigations of the DM annihilation imprint on the 21cm brightness temperature in literature that assume a redshift-dependent but spatially homogeneous boost factor calculated by integrating over analytic models for the halo mass function and halo profiles, leading to spatially uniform DM heating. While the commonly employed Zel'dovich approximation performs well on scales $k \lesssim 1 \ \text{cMpc}^{-1}$ at $z \sim 11$ (see \citealt[Figs. 2, 3]{Mesinger2011}), accurately determining the non-linear DM density field is particularly important in the context of DM annihilation which scales as $\text{d}E/\text{d}t \propto \rho_\chi^2$. Moreover, the IGM evolves decoupled from the DM and is subject to hydrodynamics in our simulations, thus causing shock heating which is often neglected when modeling the 21cm signal, although it constitutes the major heating source at high redshift \citep{Furlanetto2004a}. This makes a direct comparison of our results with approximate methods such as \textsc{21cmFAST} challenging: in our fiducial simulation without DM annihilation, shocks heat gas bound in structures to $T_\text{gas}  \sim 10^4 \ \text{K}$ despite baryonic cooling physics and the absence of astrophysical heating sources, whereas $T_\text{gas} \lesssim 5 \ \text{K}$ at $z \sim 11$ in the entire IGM using \textsc{21cmFAST} when X-ray heating from stars is deactivated. 
\par For investigating how sensitive $\delta T_b$ is with respect to the localized computation of the DM heating, we ran simulations in which we replaced the local DM density in Equation \eqref{eq:DMAF} by the average DM density $\langle \rho_\chi \rangle^2$ at each redshift and deposited the resulting amount of energy homogeneously into the IGM, mimicking a spatially unresolved treatment of DM heating characteristic for analytical models. With this simplification, the $\delta T_b$ distribution becomes slightly narrower as expected, but the difference towards the localized energy calculation is small. However, we find that the isotropic 21cm 3D power spectrum is underestimated by $7\%$ on a scale of $k = 0.1 \ \text{cMpc}^{-1}$ in comparison with localized heating for the $10 \ \text{GeV}$ case, suggesting that probing DM-baryon interactions in cosmological simulations will become increasingly important as precise 21cm measurements at high redshift become available.

\par SKA1-Low will be able to image regions of $\sim 5'$ $(10 \ h^{-1} \ \text{cMpc})$ at $z \sim 11$ at a noise level of $10 \ \text{mK}$ \citep{Mellema2015} -- below the difference between the global 21cm brightness temperatures for a $100 \ \text{GeV}$ thermal WIMP (assuming $f = 1$) annihilating via s-wave annihilation and the fiducial model without DM annihilation, which amounts to $\sim 15 \ \text{mK}$. Although disentangling DM annihilation from astrophysical signatures in the 21cm signal is a difficult undertaking, the more uniform and slower heating from DM annihilation might be distinguished from the higher heating rate of astrophysical sources by measuring the gradient $\text{d}\delta T_b/\text{d}\nu$ as proposed by \citet{Valdes2013}. Furthermore, the different shapes of the $\delta T_b$ distributions for varying annihilation strengths in Fig. \ref{fig:Tb_hist} suggest that higher moments such as the skewness, which will be probed by the SKA, may be leveraged to constrain DM heating. If the unexpectedly deep absorption trough measured by EDGES \citep{Bowman2018} in the global 21cm signal centered at $78 \ \text{MHz}$ (corresponding to $z \sim 17$) is confirmed by other instruments, this will imply competitive constraints on the DM particle, as demonstrated by \citet{DAmico2018, Liu2018}.
The EoR, Cosmic Dawn, and the Dark Ages are promising epochs to look for the effects of DM annihilation, and upcoming measurements of the global 21cm signal, the 21cm power spectrum, and in particular tomographic 21cm images from the SKA all have the potential to further constrain the nature of DM. 
Cosmological simulations will play a key role for the interpretation of the future data.  We will present further results covering a broad range of redshifts and with additional X-ray heating from stars in a subsequent publication.

\section*{Acknowledgments}
The authors acknowledge the National Computational Infrastructure (NCI), which is supported by the Australian Government, and the Pawsey Supercomputing Centre with funding from the Australian Government and the Government of Western Australia for providing services and computational resources on the supercomputers Raijin, Gadi, and Magnus. 
FL is supported by the University of Sydney International Scholarship (USydIS). PJE is funded by the Australian Research Council Centre of Excellence for All Sky Astrophysics in 3 Dimensions (ASTRO 3D), through project number CE170100013.



\begin{thebibliography}{}
\expandafter\ifx\csname natexlab\endcsname\relax\def\natexlab#1{#1}\fi
\providecommand{\url}[1]{\href{#1}{#1}}
\providecommand{\dodoi}[1]{doi:~\href{http://doi.org/#1}{\nolinkurl{#1}}}
\providecommand{\doeprint}[1]{\href{http://ascl.net/#1}{\nolinkurl{http://ascl.net/#1}}}
\providecommand{\doarXiv}[1]{\href{https://arxiv.org/abs/#1}{\nolinkurl{https://arxiv.org/abs/#1}}}

\bibitem[{Abdallah {et~al.}(2015)Abdallah, Araujo, Arbey, Ashkenazi, Belyaev,
  Berger, Boehm, Boveia, Brennan, Brooke, Buchmueller, Buckley, Busoni,
  Calibbi, Chauhan, Daci, Davies, {De Bruyn}, {De Jong}, {De Roeck}, de~Vries,
  {Del Re}, {De Simone}, {Di Simone}, Doglioni, Dolan, Dreiner, Ellis, Eno,
  Etzion, Fairbairn, Feldstein, Flaecher, Feng, Fox, Genest, Gouskos, Gramling,
  Haisch, Harnik, Hibbs, Hoh, Hopkins, Ippolito, Jacques, Kahlhoefer, Khoze,
  Kirk, Korn, Kotov, Kunori, Landsberg, Liem, Lin, Lowette, Lucas, Malgeri,
  Malik, McCabe, Mete, Morgante, Mrenna, Nakahama, Newbold, Nordstrom, Pani,
  Papucci, Pataraia, Penning, Pinna, Polesello, Racco, Re, Riotto, Rizzo,
  Salek, Sarkar, Schramm, Skubic, Slone, Smirnov, Soreq, Sumner, Tait, Thomas,
  Tomalin, Tunnell, Vichi, Volansky, Weiner, West, Wielers, Worm, Yavin,
  Zaldivar, Zhou, \& Zurek}]{Abdallah2015}
Abdallah, J., Araujo, H., Arbey, A., {et~al.} 2015, Phys. Dark Universe, 9, 8,
  \dodoi{10.1016/j.dark.2015.08.001}

\bibitem[{Ackermann {et~al.}(2015)Ackermann, Albert, Anderson, Atwood, Baldini,
  Barbiellini, Bastieri, Bechtol, Bellazzini, Bissaldi, Blandford, Bloom,
  Bonino, Bottacini, Brandt, Bregeon, Bruel, Buehler, Caliandro, Cameron,
  Caputo, Caragiulo, Caraveo, Cecchi, Charles, Chekhtman, Chiang, Chiaro,
  Ciprini, Claus, Cohen-Tanugi, Conrad, Cuoco, Cutini, D'Ammando, de~Angelis,
  de~Palma, Desiante, Digel, {Di Venere}, Drell, Drlica-Wagner, Essig, Favuzzi,
  Fegan, Ferrara, Focke, Franckowiak, Fukazawa, Funk, Fusco, Gargano,
  Gasparrini, Giglietto, Giordano, Giroletti, Glanzman, Godfrey, Gomez-Vargas,
  Grenier, Guiriec, Gustafsson, Hays, Hewitt, Horan, Jogler, J{\'{o}}hannesson,
  Kuss, Larsson, Latronico, Li, Li, {Llena Garde}, Longo, Loparco, Lubrano,
  Malyshev, Mayer, Mazziotta, McEnery, Meyer, Michelson, Mizuno, Moiseev,
  Monzani, Morselli, Murgia, Nuss, Ohsugi, Orienti, Orlando, Ormes, Paneque,
  Perkins, Pesce-Rollins, Piron, Pivato, Porter, Rain{\`{o}}, Rando, Razzano,
  Reimer, Reimer, Ritz, S{\'{a}}nchez-Conde, Schulz, Sehgal, Sgr{\`{o}},
  Siskind, Spada, Spandre, Spinelli, Strigari, Tajima, Takahashi, Thayer,
  Tibaldo, Torres, Troja, Vianello, Werner, Winer, Wood, Wood, Zaharijas, \&
  Zimmer}]{Ackermann2015}
Ackermann, M., Albert, A., Anderson, B., {et~al.} 2015, Phys. Rev. Lett., 115,
  231301, \dodoi{10.1103/PhysRevLett.115.231301}

\bibitem[{Albert {et~al.}(2017)Albert, Anderson, Bechtol, Drlica-Wagner, Meyer,
  S{\'{a}}nchez-Conde, Strigari, Wood, Abbott, Abdalla, Benoit-L{\'{e}}vy,
  Bernstein, Bernstein, Bertin, Brooks, Burke, Rosell, Kind, Carretero, Crocce,
  Cunha, D'Andrea, da~Costa, Desai, Diehl, Dietrich, Doel, Eifler, Evrard,
  Neto, Finley, Flaugher, Fosalba, Frieman, Gerdes, Goldstein, Gruen, Gruendl,
  Honscheid, James, Kent, Kuehn, Kuropatkin, Lahav, Li, Maia, March, Marshall,
  Martini, Miller, Miquel, Neilsen, Nord, Ogando, Plazas, Reil, Romer, Rykoff,
  Sanchez, Santiago, Schubnell, Sevilla-Noarbe, Smith, Soares-Santos, Sobreira,
  Suchyta, Swanson, Tarle, Vikram, Walker, \& Wechsler}]{Albert2017}
Albert, A., Anderson, B., Bechtol, K., {et~al.} 2017, ApJ, 834, 110,
  \dodoi{10.3847/1538-4357/834/2/110}

\bibitem[{Ba{\~{n}}ados {et~al.}(2018)Ba{\~{n}}ados, Venemans, Mazzucchelli,
  Farina, Walter, Wang, Decarli, Stern, Fan, Davies, Hennawi, Simcoe, Turner,
  Rix, Yang, Kelson, Rudie, \& Winters}]{Banados2018}
Ba{\~{n}}ados, E., Venemans, B.~P., Mazzucchelli, C., {et~al.} 2018, Nature,
  553, 473, \dodoi{10.1038/nature25180}

\bibitem[{Barkana \& Loeb(2005)}]{Barkana2005}
Barkana, R., \& Loeb, A. 2005, ApJ, 624, L65, \dodoi{10.1086/430599}

\bibitem[{Bertone {et~al.}(2005)Bertone, Hooper, \& Silk}]{Bertone2005}
Bertone, G., Hooper, D., \& Silk, J. 2005, Phys. Rep., 405, 279,
  \dodoi{10.1016/j.physrep.2004.08.031}

\bibitem[{Bowman {et~al.}(2018)Bowman, Rogers, Monsalve, Mozdzen, \&
  Mahesh}]{Bowman2018}
Bowman, J.~D., Rogers, A. E.~E., Monsalve, R.~A., Mozdzen, T.~J., \& Mahesh, N.
  2018, Nature, 555, 67, \dodoi{10.1038/nature25792}

\bibitem[{Chuzhoy(2008)}]{Chuzhoy2008}
Chuzhoy, L. 2008, ApJ, 679, L65, \dodoi{10.1086/589504}

\bibitem[{Cumberbatch {et~al.}(2010)Cumberbatch, Lattanzi, \&
  Silk}]{Cumberbatch2010}
Cumberbatch, D.~T., Lattanzi, M., \& Silk, J. 2010, Phys. Rev. D, 82, 1,
  \dodoi{10.1103/PhysRevD.82.103508}

\bibitem[{Cuoco {et~al.}(2017)Cuoco, Kr{\"{a}}mer, \& Korsmeier}]{Cuoco2017}
Cuoco, A., Kr{\"{a}}mer, M., \& Korsmeier, M. 2017, Phys. Rev. Lett., 118,
  191102, \dodoi{10.1103/PhysRevLett.118.191102}

\bibitem[{D'Amico {et~al.}(2018)D'Amico, Panci, \& Strumia}]{DAmico2018}
D'Amico, G., Panci, P., \& Strumia, A. 2018, Phys. Rev. Lett., 121, 011103,
  \dodoi{10.1103/PhysRevLett.121.011103}

\bibitem[{DeBoer {et~al.}(2017)DeBoer, Parsons, Aguirre, Alexander, Ali,
  Beardsley, Bernardi, Bowman, Bradley, Carilli, Cheng, Acedo, Dillon,
  Ewall-Wice, Fadana, Fagnoni, Fritz, Furlanetto, Glendenning, Greig,
  Grobbelaar, Hazelton, Hewitt, Hickish, Jacobs, Julius, Kariseb, Kohn,
  Lekalake, Liu, Loots, MacMahon, Malan, Malgas, Maree, Martinot, Mathison,
  Matsetela, Mesinger, Morales, Neben, Patra, Pieterse, Pober, Razavi-Ghods,
  Ringuette, Robnett, Rosie, Sell, Smith, Syce, Tegmark, Thyagarajan, Williams,
  \& Zheng}]{DeBoer2017}
DeBoer, D.~R., Parsons, A.~R., Aguirre, J.~E., {et~al.} 2017, Publ. Astron.
  Soc. Pacific, 129, 045001, \dodoi{10.1088/1538-3873/129/974/045001}

\bibitem[{Dewdney {et~al.}(2013)Dewdney, Turner, Millenaar, Mccool, Lazio, \&
  Cornwell}]{Dewdney2013}
Dewdney, P.~E., Turner, W., Millenaar, R., {et~al.} 2013, SKA Technical
  Document

\bibitem[{Elahi {et~al.}(2019)Elahi, Ca{\~{n}}as, Poulton, Tobar, Willis,
  Lagos, Power, \& Robotham}]{Elahi2019}
Elahi, P.~J., Ca{\~{n}}as, R., Poulton, R. J.~J., {et~al.} 2019, PASA, 36,
  e021, \dodoi{10.1017/pasa.2019.12}

\bibitem[{Evoli {et~al.}(2014)Evoli, Mesinger, \& Ferrara}]{Evoli2014}
Evoli, C., Mesinger, A., \& Ferrara, A. 2014, JCAP, 2014,
  \dodoi{10.1088/1475-7516/2014/11/024}

\bibitem[{Evoli {et~al.}(2012)Evoli, Vald{\'{e}}s, Ferrara, \&
  Yoshida}]{Evoli2012}
Evoli, C., Vald{\'{e}}s, M., Ferrara, A., \& Yoshida, N. 2012, MNRAS, 422, 420,
  \dodoi{10.1111/j.1365-2966.2012.20624.x}

\bibitem[{Field(1958)}]{Field1958}
Field, G.~B. 1958, Proceedings of the IRE, 46, 240

\bibitem[{Furlanetto \& Loeb(2004)}]{Furlanetto2004a}
Furlanetto, S.~R., \& Loeb, A. 2004, ApJ, 611, 642, \dodoi{10.1086/422242}

\bibitem[{Furlanetto {et~al.}(2006{\natexlab{a}})Furlanetto, Oh, \&
  Pierpaoli}]{Furlanetto2006a}
Furlanetto, S.~R., Oh, S.~P., \& Pierpaoli, E. 2006{\natexlab{a}}, Phys. Rev.
  D, 74, 103502, \dodoi{10.1103/PhysRevD.74.103502}

\bibitem[{Furlanetto {et~al.}(2006{\natexlab{b}})Furlanetto, {Peng Oh}, \&
  Briggs}]{Furlanetto2006}
Furlanetto, S.~R., {Peng Oh}, S., \& Briggs, F.~H. 2006{\natexlab{b}}, Phys.
  Rep., 433, 181, \dodoi{10.1016/j.physrep.2006.08.002}

\bibitem[{Hopkins(2015)}]{GIZMO}
Hopkins, P.~F. 2015, MNRAS, 450, 53, \dodoi{10.1093/mnras/stv195}

\bibitem[{Hopkins {et~al.}(2018)Hopkins, Wetzel, Kere{\v{s}},
  Faucher-Gigu{\`{e}}re, Quataert, Boylan-Kolchin, Murray, Hayward,
  Garrison-Kimmel, Hummels, Feldmann, Torrey, Ma, Angl{\'{e}}s-Alc{\'{a}}zar,
  Su, Orr, Schmitz, Escala, Sanderson, Grudi{\'{c}}, Hafen, Kim, Fitts,
  Bullock, Wheeler, Chan, Elbert, \& Narayanan}]{FIRE2}
Hopkins, P.~F., Wetzel, A., Kere{\v{s}}, D., {et~al.} 2018, MNRAS, 480, 800,
  \dodoi{10.1093/mnras/sty1690}

\bibitem[{Iwanus {et~al.}(2017)Iwanus, Elahi, \& Lewis}]{Iwanus2017}
Iwanus, N., Elahi, P.~J., \& Lewis, G.~F. 2017, MNRAS, 472, 1214,
  \dodoi{10.1093/mnras/stx1974}

\bibitem[{Iwanus {et~al.}(2019)Iwanus, Elahi, List, \& Lewis}]{Iwanus2019}
Iwanus, N., Elahi, P.~J., List, F., \& Lewis, G.~F. 2019, MNRAS, 485, 1420,
  \dodoi{10.1093/mnras/stz435}

\bibitem[{Leane {et~al.}(2018)Leane, Slatyer, Beacom, \& Ng}]{Leane2018}
Leane, R.~K., Slatyer, T.~R., Beacom, J.~F., \& Ng, K. C.~Y. 2018, Phys. Rev.
  D, 98, 23016, \dodoi{10.1103/PhysRevD.98.023016}

\bibitem[{List {et~al.}(2019)List, Iwanus, Elahi, \& Lewis}]{List2019}
List, F., Iwanus, N., Elahi, P.~J., \& Lewis, G.~F. 2019, MNRAS, 489, 4217,
  \dodoi{10.1093/mnras/stz2287}

\bibitem[{Liu {et~al.}(2020)Liu, Ridgway, \& Slatyer}]{Liu2020}
Liu, H., Ridgway, G.~W., \& Slatyer, T.~R. 2020, Phys. Rev. D, 101, 023530,
  \dodoi{10.1103/PhysRevD.101.023530}

\bibitem[{Liu \& Slatyer(2018)}]{Liu2018}
Liu, H., \& Slatyer, T.~R. 2018, Phys. Rev. D, 98, 023501,
  \dodoi{10.1103/PhysRevD.98.023501}

\bibitem[{Liu {et~al.}(2016)Liu, Slatyer, \& Zavala}]{Liu2016}
Liu, H., Slatyer, T.~R., \& Zavala, J. 2016, Phys. Rev. D, 94, 063507,
  \dodoi{10.1103/PhysRevD.94.063507}

\bibitem[{Lopez-Honorez {et~al.}(2016)Lopez-Honorez, Mena, Molin{\'{e}},
  Palomares-Ruiz, \& Vincent}]{Lopez-Honorez2016}
Lopez-Honorez, L., Mena, O., Molin{\'{e}}, {\'{A}}., Palomares-Ruiz, S., \&
  Vincent, A.~C. 2016, JCAP, 2016, \dodoi{10.1088/1475-7516/2016/08/004}

\bibitem[{Lopez-Honorez {et~al.}(2013)Lopez-Honorez, Mena, Palomares-Ruiz, \&
  Vincent}]{Lopez-Honorez2013}
Lopez-Honorez, L., Mena, O., Palomares-Ruiz, S., \& Vincent, A.~C. 2013, JCAP,
  2013, \dodoi{10.1088/1475-7516/2013/07/046}

\bibitem[{Mellema {et~al.}(2015)Mellema, Koopmans, Shukla, Datta, Mesinger, \&
  Majumdar}]{Mellema2015}
Mellema, G., Koopmans, L., Shukla, H., {et~al.} 2015, Proceedings of Advancing
  Astrophysics with the Square Kilometre Array — PoS(AASKA14), 215, 010,
  \dodoi{10.22323/1.215.0010}

\bibitem[{Mellema {et~al.}(2013)Mellema, Koopmans, Abdalla, Bernardi, Ciardi,
  Daiboo, de~Bruyn, Datta, Falcke, Ferrara, Iliev, Iocco, Jeli{\'{c}}, Jensen,
  Joseph, Labroupoulos, Meiksin, Mesinger, Offringa, Pandey, Pritchard, Santos,
  Schwarz, Semelin, Vedantham, Yatawatta, \& Zaroubi}]{Mellema2013}
Mellema, G., Koopmans, L. V.~E., Abdalla, F.~A., {et~al.} 2013, Experimental
  Astronomy, 36, 235, \dodoi{10.1007/s10686-013-9334-5}

\bibitem[{Mesinger {et~al.}(2011)Mesinger, Furlanetto, \& Cen}]{Mesinger2011}
Mesinger, A., Furlanetto, S., \& Cen, R. 2011, MNRAS, 411, 955,
  \dodoi{10.1111/j.1365-2966.2010.17731.x}

\bibitem[{Navarro {et~al.}(1997)Navarro, Frenk, \& White}]{Navarro1997}
Navarro, J.~F., Frenk, C.~S., \& White, S. D.~M. 1997, ApJ, 490, 493,
  \dodoi{10.1086/304888}

\bibitem[{Oldengott {et~al.}(2016)Oldengott, Boriero, \&
  Schwarz}]{Oldengott2016}
Oldengott, I.~M., Boriero, D., \& Schwarz, D.~J. 2016, JCAP, 2016,
  \dodoi{10.1088/1475-7516/2016/08/054}

\bibitem[{{Planck Collaboration}(2016)}]{Ade2016}
{Planck Collaboration}. 2016, A{\&}A, 594, A13,
  \dodoi{10.1051/0004-6361/201525830}

\bibitem[{{Planck Collaboration}(2018)}]{PlanckCollaboration2018}
---. 2018, preprint (arXiv:1807.06209).
\newblock \doarXiv{1807.06209}

\bibitem[{Poulin {et~al.}(2015)Poulin, Serpico, \& Lesgourgues}]{Poulin2015}
Poulin, V., Serpico, P.~D., \& Lesgourgues, J. 2015, JCAP, 2015, 041,
  \dodoi{10.1088/1475-7516/2015/12/041}

\bibitem[{Press \& Schechter(1974)}]{press1974formation}
Press, W.~H., \& Schechter, P. 1974, ApJ, 187, 425, \dodoi{10.1086/152650}

\bibitem[{Pritchard \& Loeb(2012)}]{Pritchard2012}
Pritchard, J.~R., \& Loeb, A. 2012, Reports on Progress in Physics, 75, 086901,
  \dodoi{10.1088/0034-4885/75/8/086901}

\bibitem[{Schumann(2019)}]{Schumann2019}
Schumann, M. 2019, Journal of Physics G: Nuclear and Particle Physics, 46,
  103003, \dodoi{10.1088/1361-6471/ab2ea5}

\bibitem[{Seager {et~al.}(1999)Seager, Sasselov, \& Scott}]{Seager1999}
Seager, S., Sasselov, D.~D., \& Scott, D. 1999, ApJ, 523, L1,
  \dodoi{10.1086/312250}

\bibitem[{Sheth \& Tormen(1999)}]{Sheth1999}
Sheth, R.~K., \& Tormen, G. 1999, MNRAS, 308, 119,
  \dodoi{10.1046/j.1365-8711.1999.02692.x}

\bibitem[{Slatyer(2013)}]{Slatyer2013}
Slatyer, T.~R. 2013, Phys. Rev. D, 87, 123513,
  \dodoi{10.1103/PhysRevD.87.123513}

\bibitem[{Slatyer(2016)}]{Slatyer2016}
---. 2016, Phys. Rev. D, 93, 023521, \dodoi{10.1103/PhysRevD.93.023521}

\bibitem[{Springel(2005)}]{Gadget2}
Springel, V. 2005, MNRAS, 364, 1105, \dodoi{10.1111/j.1365-2966.2005.09655.x}

\bibitem[{Springel(2015)}]{NGenic}
---. 2015, {N-GenIC: Cosmological structure initial conditions}, Astrophysics
  Source Code Library ascl:1502.003

\bibitem[{Springel \& Hernquist(2003)}]{Springel2003}
Springel, V., \& Hernquist, L. 2003, MNRAS, 339, 289,
  \dodoi{10.1046/j.1365-8711.2003.06206.x}

\bibitem[{Steigman {et~al.}(2012)Steigman, Dasgupta, \& Beacom}]{Steigman2012}
Steigman, G., Dasgupta, B., \& Beacom, J.~F. 2012, Phys. Rev. D, 86, 023506,
  \dodoi{10.1103/PhysRevD.86.023506}

\bibitem[{Vald{\'{e}}s {et~al.}(2013)Vald{\'{e}}s, Evoli, Mesinger, Ferrara, \&
  Yoshida}]{Valdes2013}
Vald{\'{e}}s, M., Evoli, C., Mesinger, A., Ferrara, A., \& Yoshida, N. 2013,
  MNRAS, 429, 1705, \dodoi{10.1093/mnras/sts458}

\bibitem[{Warren {et~al.}(2006)Warren, Abazajian, Holz, \&
  Teodoro}]{Warren2006}
Warren, M.~S., Abazajian, K., Holz, D.~E., \& Teodoro, L. 2006, ApJ, 646, 881,
  \dodoi{10.1086/504962}

\bibitem[{Wiersma {et~al.}(2009)Wiersma, Schaye, \& Smith}]{Wiersma2009}
Wiersma, R.~P., Schaye, J., \& Smith, B.~D. 2009, MNRAS, 393, 99,
  \dodoi{10.1111/j.1365-2966.2008.14191.x}

\bibitem[{Wouthuysen(1952)}]{Wouthuysen1952}
Wouthuysen, S.~A. 1952, The Astronomical Journal, 57, 31,
  \dodoi{10.1086/106661}

\bibitem[{Zel'dovich(1970)}]{zel1970gravitational}
Zel'dovich, Y.~B. 1970, A{\&}A, 5, 84

\end{thebibliography}



\end{document}